\PassOptionsToPackage{unicode}{hyperref}
\PassOptionsToPackage{hyphens}{url}
\PassOptionsToPackage{dvipsnames,svgnames,x11names}{xcolor}
\documentclass[
]{article}
\usepackage{amsmath,amssymb}
\usepackage{lmodern}
\usepackage{iftex}
\ifPDFTeX
  \usepackage[T1]{fontenc}
  \usepackage[utf8]{inputenc}
  \usepackage{textcomp} 
\else 
  \usepackage{unicode-math}
  \defaultfontfeatures{Scale=MatchLowercase}
  \defaultfontfeatures[\rmfamily]{Ligatures=TeX,Scale=1}
\fi
\IfFileExists{upquote.sty}{\usepackage{upquote}}{}
\IfFileExists{microtype.sty}{
  \usepackage[]{microtype}
  \UseMicrotypeSet[protrusion]{basicmath} 
}{}
\makeatletter
\@ifundefined{KOMAClassName}{
  \IfFileExists{parskip.sty}{%
    \usepackage{parskip}
  }{
    \setlength{\parindent}{0pt}
    \setlength{\parskip}{6pt plus 2pt minus 1pt}}
}{
  \KOMAoptions{parskip=half}}
\makeatother
\usepackage{xcolor}
\NewDocumentCommand\citeproctext{}{}
\NewDocumentCommand\citeproc{mm}{%
  \begingroup\def\citeproctext{#2}\cite{#1}\endgroup}
\makeatletter
 \let\@cite@ofmt\@firstofone
 \def\@biblabel#1{}
 \def\@cite#1#2{{#1\if@tempswa , #2\fi}}
\makeatother
\newlength{\cslhangindent}
\newlength{\csllabelwidth}
\newenvironment{CSLReferences}[2] 
 {\begin{list}{}{%
  \setlength{\itemindent}{0pt}
  \setlength{\leftmargin}{0pt}
  \setlength{\parsep}{0pt}
  \ifodd #1
   \setlength{\leftmargin}{\cslhangindent}
   \setlength{\itemindent}{-1\cslhangindent}
  \fi
  \setlength{\itemsep}{#2\baselineskip}}}
 {\end{list}}
\usepackage{calc}

\ifLuaTeX
\usepackage[bidi=basic]{babel}
\else
\usepackage[bidi=default]{babel}
\fi
\babelprovide[main,import]{american}

\def\languageshorthands#1{}
\ifLuaTeX
  \usepackage{selnolig}  
\fi
\IfFileExists{bookmark.sty}{\usepackage{bookmark}}{\usepackage{hyperref}}
\IfFileExists{xurl.sty}{\usepackage{xurl}}{} 
\hypersetup{
  pdftitle={orbitize! v4: Joint Orbit Fitting with Gaia DR4 Astrometry
and Flexible Analytic Linear Marginalization},
  pdfauthor={Clarissa R. Do Ó, Eshel Dror, Sarah Blunt, Vanessa P.
Bailey, William O. Balmer, Beth A. Biller, Quinton Blackston, Aniruddh
Chalagulla, Robert J. De Rosa, Chih-Chun Hsu, Isabela Huckabee, Farrah
Molina, Niklas Naworal, Eric L. Nielsen, Tomas Stolker, Rifah Tasnim,
David Trevascus, Edward Turner, Vidya Venkatesan, Jason J. Wang, Marie
Ygouf, Jingwen Zhang},
  pdflang={en-US},
  colorlinks=true,
  linkcolor={Maroon},
  filecolor={Maroon},
  citecolor={Blue},
  urlcolor={Blue},
  pdfcreator={LaTeX via pandoc}}

\title{orbitize! v4: Joint Orbit Fitting with Gaia DR4 Astrometry and
Flexible Analytic Linear Marginalization}

\definecolor{c53baa1}{RGB}{83,186,161}
\definecolor{c202826}{RGB}{32,40,38}

\usepackage[affil-it]{authblk}
\usepackage{orcidlink}
\author[1%
  \ensuremath\mathparagraph]{Clarissa R. Do Ó%
    \,\orcidlink{0000-0001-5173-2947}\,%
    }
\author[2%
  ]{Eshel Dror%
    }
\author[3,2%
  ]{Sarah Blunt%
    \,\orcidlink{0000-0002-3199-2888}\,%
    }
\author[4%
  ]{Vanessa P. Bailey%
    \,\orcidlink{0000-0002-5407-2806}\,%
    }
\author[5%
  ]{William O. Balmer%
    \,\orcidlink{0000-0001-6396-8439}\,%
    }
\author[6%
  ]{Beth A. Biller%
    \,\orcidlink{0000-0003-4614-7035}\,%
    }
\author[2%
  ]{Quinton Blackston%
    }
\author[2%
  ]{Aniruddh Chalagulla%
    }
\author[7%
  ]{Robert J. De Rosa%
    \,\orcidlink{0000-0002-4918-0247}\,%
    }
\author[5%
  ]{Chih-Chun Hsu%
    \,\orcidlink{0000-0002-5370-7494}\,%
    }
\author[8%
  ]{Isabela Huckabee%
    }
\author[5,9%
  ]{Farrah Molina%
    }
\author[2%
  ]{Niklas Naworal%
    }
\author[10%
  ]{Eric L. Nielsen%
    }
\author[11%
  ]{Tomas Stolker%
    \,\orcidlink{0000-0002-5823-3072}\,%
    }
\author[8%
  ]{Rifah Tasnim%
    }
\author[12%
  ]{David Trevascus%
    }
\author[6%
  ]{Edward Turner%
    }
\author[13%
  ]{Vidya Venkatesan%
    }
\author[5,15%
  ]{Jason J. Wang%
    \,\orcidlink{0000-0003-0774-6502}\,%
    }
\author[4%
  ]{Marie Ygouf%
    }
\author[14%
  ]{Jingwen Zhang%
    \,\orcidlink{0000-0002-2696-2406}\,%
    }

\affil[1]{Department of Astronomy, California Institute of Technology,
Pasadena, CA, USA%
  }
\affil[2]{Department of Astronomy and Astrophysics, University of
California, Santa Cruz, Santa Cruz, CA, USA%
  }
\affil[3]{Department of Astrophysical and Planetary Sciences, University
of Colorado Boulder, Boulder, CO, USA%
  }
\affil[4]{Jet Propulsion Laboratory, California Institute of Technology,
Pasadena, CA, USA%
  }
\affil[5]{Center for Interdisciplinary Exploration and Research in
Astrophysics (CIERA), Northwestern University, Evanston, IL, USA%
  }
\affil[6]{Institute for Astronomy, School of Physics and Astronomy,
University of Edinburgh, Edinburgh, UK%
  }
\affil[7]{European Southern Observatory, Santiago, Chile%
  }
\affil[8]{Department of Astronomy and Carl Sagan Institute, Cornell
University, Ithaca, NY, USA%
  }
\affil[9]{Department of Physics and Engineering, University of
St.~Thomas, Houston, TX, USA%
  }
\affil[10]{Department of Astronomy, New Mexico State University, Las
Cruces, NM, USA%
  }
\affil[11]{Leiden Observatory, Leiden University, Leiden, The
Netherlands%
  }
\affil[12]{Max Planck Institute for Astronomy, Heidelberg, Germany%
  }
\affil[13]{University of California, Riverside, Riverside, CA, USA%
  }
\affil[14]{Department of Physics, University of California, Santa
Barbara, Santa Barbara, CA, USA%
  }
\affil[15]{Department of Physics and Astronomy Northwestern University,
Evanston, IL 60208, USA%
  }
\affil[$\mathparagraph$]{Corresponding author: %
}
\date{19 November 2026}

\begin{document}
\maketitle

\section{Summary}\label{summary}

\texttt{orbitize!} is a Python package for Bayesian orbit fitting of
planetary and stellar companions (\citeproc{ref-Blunt:2020}{Blunt et
al., 2020}). It provides a unified framework for fitting relative
astrometry, radial velocities, and absolute astrometry, supporting
analyses of both resolved and unresolved systems. Available posterior
sampling methods include parallel-tempered Markov chain Monte Carlo
(MCMC) through \texttt{ptemcee} (\citeproc{ref-emcee}{Foreman-Mackey et
al., 2013}; \citeproc{ref-ptemcee}{Vousden et al., 2016}) and nested
sampling through \texttt{dynesty} (\citeproc{ref-Speagle:2020}{Speagle,
2020}), \texttt{nautilus} (\citeproc{ref-nautilus}{Lange, 2023}), and
\texttt{pymultinest} (\citeproc{ref-pymultinest}{Buchner et al., 2014};
\citeproc{ref-multinest}{Feroz et al., 2019}). This paper accompanies
the release of \texttt{orbitize!} version 4.0 and describes the
functionality added since version 3.0
(\citeproc{ref-2024JOSS....9.6756B}{Blunt et al., 2024}), with a
particular focus on joint orbit fitting with Gaia DR4 epoch astrometry
and the implementation of the Orbits for the Extremely Impatient (OFTeI)
algorithm (Blunt et al.~2026b, in preparation).

\section{Statement of need}\label{statement-of-need}

The fourth data release (DR4) of the \emph{Gaia} mission
(\citeproc{ref-2016Aux26A...595A...1G}{Gaia Collaboration et al., 2016})
will offer new opportunities to measure companion masses and
characterize planetary system architectures through epoch astrometry:.
the individual measurements of a source's position along Gaia's scan
direction at different observation times. These measurements will offer
new opportunities to characterize planetary system architectures. Many
companions detected through Gaia astrometry are also expected to be
detectable with other methods, particularly radial velocities
(\citeproc{ref-LammersWinn:2026}{Lammers \& Winn, 2026};
\citeproc{ref-Winn:2022}{Winn, 2022}). Combining astrometry with stellar
radial velocities can break the mass--inclination degeneracy of RV-only
fits and constrain three-dimensional orbital orientations. For directly
imaged systems, relative astrometry can only constrain the total system
mass. Adding the primary's absolute orbital motion can constrain the
individual component masses.

\texttt{orbitize!} v4 provides these capabilities in an open source
framework, by modeling individual Gaia along-scan measurements and
fitting them alone or jointly with stellar radial velocities, companion
radial velocities, and relative astrometry. Additionally, implementation
of the Orbits for the Extremely Impatient Algorithm, a novel analytic
marginalization scheme for relative and absolute astrometry (Blunt et
al.~2026b, in prep), together with an optimized Kepler solver, enables
much faster convergence relative to earlier versions of the code.

\subsection{Gaia DR4 epoch astrometry}\label{gaia-dr4-epoch-astrometry}

Version 4 introduces the \texttt{gaia.DR4LogProb} class for fitting Gaia
epoch astrometry. The forward model includes five astrometric
parameters: two position offsets, \(\alpha_{*,0}\) and \(\delta_0\), two
proper motion components, \(\mu_{\alpha*}\) and \(\mu_\delta\), and
parallax, \(\varpi\). Together with the primary's orbital reflex motion
about the system barycenter, these describe its apparent motion on the
sky (\citeproc{ref-Brandt:2024}{Brandt, 2024};
\citeproc{ref-Nielsen:2020}{Nielsen et al., 2020}). Users can choose
among the orbital fitting bases supported by \texttt{orbitize!} (see
\citeproc{ref-Blunt:2020}{Blunt et al., 2020};
\citeproc{ref-Blunt:2023a}{Blunt et al., 2023}, for orbital parameter
definitions). The predicted sky motion is then projected onto the scan
direction at each Gaia observation. For measurement \(i\), the predicted
along-scan position is

\[
\begin{aligned}
\eta_i ={}&
\left(\alpha_{*,0} + \mu_{\alpha*}\Delta t_i
      + x_{\star,i}\right)\sin\psi_i \\
&+
\left(\delta_0 + \mu_{\delta}\Delta t_i
      + y_{\star,i}\right)\cos\psi_i
+ \varpi \Pi_i ,
\end{aligned}
\]

where \(\psi_i\) is the scan position angle, \(\Pi_i\) is the along-scan
parallax factor, and \(\varpi\) is the parallax. The proper motion
components are \(\mu_{\alpha*} = \dot{\alpha}\cos\delta\) in right
ascension and \(\mu_\delta = \dot{\delta}\) in declination, both
expressed in mas\,jyr\(^{-1}\). The position offsets \(\alpha_{*,0}\) and
\(\delta_0\) are defined at a reference epoch (by default the Gaia DR4
reference epoch of
\href{https://esa.github.io/gaia-supdate/gaiasupdate.html\#gaiasupdate.constants.DR4_REFERENCE_EPOCH}{J2017.5}),
and \(\Delta t_i\) is the elapsed time in Julian years. The quantities
\(x_{\star,i}\) and \(y_{\star,i}\) are the photocenter's barycentric
orbital offsets in right ascension, including the \(\cos\delta\) factor,
and declination, respectively. For systems with multiple companions,
these offsets include their combined reflex signal, and orbital
parameters are defined in Jacobi coordinates
(\citeproc{ref-Lacour:2021}{Lacour et al., 2021}).

The astrometric log-likelihood is

\[
\ln \mathcal{L}_{\mathrm{DR4}}
=
-\frac{1}{2}\sum_i
\left[
\frac{(d_i-\eta_i)^2}{\sigma_i^2+s_{\mathrm{G}}^2}
+
\ln\!\left(2\pi[\sigma_i^2+s_{\mathrm{G}}^2]\right)
\right],
\]

where \(d_i\) and \(\sigma_i\) are the measured along-scan position and
its reported uncertainty. \(\eta_i\) denotes the predicted along-scan
angular offset for observation \(i\) in milliarcseconds. An additional
jitter term, \(s_{\mathrm{G}}\), can be fixed or fitted and is added in
quadrature to the reported uncertainties, as in Holl et al.
(\citeproc{ref-2023Aux26A...674A..10H}{2023}). Position offsets and
proper motions are sampled explicitly alongside the orbital parameters
and parallax with configurable priors.

Gaia measurements can be fitted alone or combined with the primary
radial velocities, companion radial velocities, relative astrometry, and
the Hipparcos IAD (\citeproc{ref-Nielsen:2020}{Nielsen et al., 2020}).
All observations are predicted from a shared orbital model, and since we
assume the datasets are independent, we add the log-likelihoods for all
data. Priors on shared parameters enter the posterior once. The
resulting posterior is sampled with the user's choice of posterior
sampling algorithm (e.g.~MCMC, nested sampling), allowing joint
constraints on orbital and astrometric parameters. \texttt{orbitize!}
automatically sets up the parameters required by the supplied datasets,
and users can select the orbital fitting basis and customize the priors
before sampling.

\section{Major features added since
v3}\label{major-features-added-since-v3}

For a detailed overview of the \texttt{orbitize!} API and core
functionality, we refer the reader to Blunt et al.
(\citeproc{ref-Blunt:2020}{2020}). This section highlights the major
features added since version 3.0. For a complete list of changes, see
the
\href{https://orbitize.readthedocs.io/en/latest/\#changelog}{changelog}.
Major new features since v3 include:

\begin{enumerate}
\def\labelenumi{\arabic{enumi}.}
\item
  A likelihood for Gaia DR4 epoch astrometry, implemented in
  \texttt{gaia.DR4LogProb}. The model projects the primary's motion onto
  each observation's scan direction. An optional astrometric jitter
  term, either fixed or fitted, is added in quadrature to the reported
  measurement uncertainties. Gaia measurements can be fitted alone or
  jointly with other datasets.
\item
  The \texttt{simulate\_dr4} module generates mock Gaia DR4 along-scan
  measurements from a specified orbital solution and
  \href{https://gaia.esac.esa.int/gost/}{Gaia Observation Forecast Tool
  (GOST)} scanning predictions. The simulations include stellar reflex
  motion, proper motion, parallax, and Gaussian noise drawn using a
  magnitude-dependent uncertainty model adopted from
  \href{https://github.com/kareemelbadry/gaiamock}{Gaiamock}
  (\citeproc{ref-ElBadry:2024}{El-Badry et al., 2024}), based on the
  precision curves presented by Holl et al.
  (\citeproc{ref-2023Aux26A...674A..10H}{2023}). The resulting tables
  can be passed directly to the DR4 likelihood for injection recovery
  tests.
\item
  Worked examples demonstrate the new fitting and plotting capabilities.
  The
  \href{https://orbitize.readthedocs.io/en/latest/tutorials/Gaia_DR4_BH3.html}{Gaia
  BH3 Tutorial} demonstrates fitting Gaia epoch astrometry alone for the
  black hole binary discovered by Gaia Collaboration et al.
  (\citeproc{ref-2024Aux26A...686L...2G}{2024}). The
  \href{https://orbitize.readthedocs.io/en/latest/tutorials/Gaia_DR4_Gaia4.html}{Gaia-4
  Tutorial} demonstrates a joint fit with stellar radial velocities from
  Stefánsson et al. (\citeproc{ref-2025AJ....169..107S}{2025}), while
  the
  \href{https://orbitize.readthedocs.io/en/latest/tutorials/Gaia_DR4_EAS.html}{HD
  114762 Tutorial} provides another example of joint RV and Gaia epoch
  astrometry fitting. These three tutorials use epoch astrometry from
  the \href{https://www.cosmos.esa.int/web/gaia/dr4-prerelease}{June
  2026 Gaia DR4 prerelease}. The
  \href{https://orbitize.readthedocs.io/en/latest/tutorials/HD160691_Joint_Fit.html}{\(\mu\)
  Ara Tutorial} demonstrates generating mock Gaia DR4 astrometry for a
  four-planet system and jointly fitting it with observed stellar radial
  velocities. All four examples illustrate posterior sampling and
  visualization of the inferred orbits.
\item
  The \texttt{marginalizer} module provides a framework for
  implementation of analytic marginalization schemes in
  \texttt{orbitize!}. The Orbits for the Extremely Impatient (OFTeI)
  algorithm, implemented in \texttt{marginalizer.OFTeI\_seppa}, allows
  partial linear marginalization over the semimajor axis and position
  angle of nodes for systems with one secondary and relative astrometry
  data. Similarly, the algorithm implemented in
  \texttt{marginalizer.OFTeI\_DR4\_gaia} performs partial linear
  marginalization over proper motion, position offset, parallax,
  semimajor axis, and position angle of nodes for systems with one
  secondary and astrometric data from Gaia DR4.
\item
  The optimization of the \texttt{kepler} module significantly speeds up
  the solving of kepler's equations by migrating more of the code to C.
  The improvement is most significant in evaluating a single set of
  parameters. For samplers such as Markov Chain Monte Carlo which
  evaluate one set of parameters at a time, the optimized sampler is
  12.39 times faster, with the calculation done by the \texttt{kepler}
  module taking about 2\% of the previous time. For evaluating an array
  of 30,000 sets of parameters, the optimized calculation takes about
  65\% of the previous time.
\item
  Refactoring of the functions in the \texttt{plot} module into the
  \texttt{plot.Plotter} class facilitates clearer code flow and new
  plotting capabilities. The \texttt{plot.Plotter} class precomputes
  shared attributes of the different plotting methods when initialized,
  so that they are not recomputed on each call of a plotting method. The
  \texttt{plot.Plotter.plot\_orbits} method has new capabilities for
  plotting the orbits of multiple bodies, displaying correlated
  errorbars on right ascension/declination data, and customizing
  grouping of sources of error for radial velocity.
  \texttt{plot.Plotter.plot\_residuals} similarly supports multiple
  bodies and the separation of errors. Multiple tutorials, especially
  the
  \href{https://github.com/sblunt/orbitize/blob/v4/docs/tutorials/Plotting_tutorial.ipynb}{Advanced
  Plotting Tutorial}, demonstrate new plotting capabilities and many new
  customization options.
\item
  Joint fitting for reflected-light photometry with Lambertian disk
  approximation, following the implementation outlined in
  (\citeproc{ref-2026RNAAS..10..203M}{Molina et al., 2026}). This
  implementation adopts relative brightness
  \(\mathrm{(brightness)} \propto \frac{A\alpha}{R^2}\) where \(A\) is
  Albedo; \(\alpha\) is the angle between the star, the planet, and the
  observer; and \(R\) is the distance between the planet and star.
  \footnote{In (\citeproc{ref-2026RNAAS..10..203M}{Molina et al.,
    2026}), equation A5 states that R is the planetary radius, which in
    this context means the distance between the planet and the star (see
    their equation A1)}
\item
  Support for two importance nested sampling algorithms. The sampler
  implemented in \texttt{sampler.NautilusSampler} uses the
  \texttt{Nautilus-Sampler} package and determines the boundaries of
  nested sampling shells using deep learning
  (\citeproc{ref-nautilus}{Lange, 2023}). The \texttt{sampler.MultiNest}
  sampler uses the \texttt{pymultinest} package
  (\citeproc{ref-pymultinest}{Buchner et al., 2014};
  \citeproc{ref-multinest}{Feroz et al., 2019}).
\end{enumerate}

\section{Verification and
Documentation}\label{verification-and-documentation}

For details of the automated testing and documentation infrastructure,
see the ``Verification and Documentation'' section of Blunt et al.
(\citeproc{ref-2024JOSS....9.6756B}{2024}).

\section{Comparison to Similar
Packages}\label{comparison-to-similar-packages}

For an overview of related orbit fitting packages, we refer readers to
Blunt et al. (\citeproc{ref-2024JOSS....9.6756B}{2024}). Current
capabilities and examples for \texttt{orbitize!} are described in our
\href{https://orbitize.readthedocs.io/en/latest/}{documentation}.
Different orbit fitting packages offer different approaches to model
construction and posterior inference. \texttt{orbitize!} v4 introduces a
generic framework for marginalizing over linear model parameters,
reducing the dimensionality of the sampled parameter space.
\texttt{kima} (\citeproc{ref-Baycroft:2026}{Baycroft et al., 2026})
provides the \texttt{GAIAmodel} and \texttt{RVGAIAmodel} for fitting
Gaia epoch astrometry alone or jointly with radial velocities, using
diffusive nested sampling. \texttt{orvara} (\citeproc{ref-orvara}{Brandt
et al., 2021}) analytically marginalizes over parallax, barycentric
proper motion, and instrument-specific radial velocity zero points.
\texttt{octofitter} (\citeproc{ref-Thompson:2023}{Thompson et al.,
2023}) supports gradient-based inference using the No-U-Turn variant of
Hamiltonian Monte Carlo and can also fit Gaia DR4 epoch astrometry, as
illustrated in its
\href{https://sefffal.github.io/Octofitter.jl/stable/gaia-dr4-prerelease/}{prerelease-data
tutorial}, which includes Gaia BH3. The \texttt{exoplanet} toolkit
(\citeproc{ref-exoplanet}{Foreman-Mackey et al., 2021}) provides tools
for building and fitting models of transit and radial velocity data
using the Bayesian modeling package \texttt{PyMC}. \texttt{ExoPlanetLab}
(\citeproc{ref-Hobbs:2026}{Hobbs, 2026}) also supports joint modeling of
Gaia epoch astrometry, radial velocities, and relative astrometry, with
multiple posterior sampling methods and a Student-\(t\) likelihood. A
detailed performance comparison between \texttt{orbitize!} v4 and
similar packages will be presented in a forthcoming paper (Dror et al.,
in preparation).

\section{Acknowledgements}\label{acknowledgements}

C.D.O acknowledges support from the B. Thomas Soifer Fellowship at the
California Institute of Technology.

This paper describes additions made to \texttt{orbitize!} between
versions 3.0.0 and 4.0.0. We are extremely grateful to Isabel Angelo,
Henry Ngo, James Graham, Logan Pearce, and Malena Rice, who contribtued
code to version 1.0.0, and are included as authors in Blunt et al.
(\citeproc{ref-Blunt:2020}{2020}).

We also thank Lea Hirsch, Roberto Tejada, Vighnesh Nagpal, Tirth
Dharmesh Surti, Sofia Covarrubias, Thea McKenna, Rodrigo Ferrer Chávez,
Jorge Llop-Sayson, Mireya Arora, Amanda Chavez, Devin Cody, Saanika
Choudhary, Adam J. R. W. Smith, and Hannah Gallamore for their
contributions to \texttt{orbitize!} v3
(\citeproc{ref-2024JOSS....9.6756B}{Blunt et al., 2024}).

We are also grateful for useful discussions with William Thompson.

\section*{References}\label{references}
\addcontentsline{toc}{section}{References}

\protect\phantomsection\label{refs}

\end{document}